\documentclass[prb,nofootinbib,twocolumn]{revtex4} 

\usepackage{graphicx}
\usepackage{dcolumn}
\usepackage{bm}
\usepackage{threeparttable}
\usepackage{times}
\usepackage{mathptmx}
\usepackage{lscape}
\usepackage{natbib}
\usepackage{amsmath}
\usepackage{amssymb}
\usepackage{braket}
\usepackage{comment}
\usepackage{color}

\def\degree{\kern-.2em\r{}\kern-.3em}

\begin{document}

\title{ Exact Transport-Information Correspondence for Generalized Discretization Schemes  }

\author{Koretaka Yuge$^{\dagger}$}
\affiliation{
Department of Materials Science and Engineering,  Kyoto University, Sakyo, Kyoto 606-8501, Japan\\
}%

\begin{abstract}
{ 
Discretization of continuous probability distributions is an operation on the sample space, and the structure induced by this operation cannot generally be determined solely from information geometry. In our recent work on unavoidable canonical nonlinearity (UCN), we established a non-arbitrary connection between the transport cost associated with discretization and the information-geometric indistinguishability of continuous probability distributions. Through this transport-information correspondence, UCN quantifies the information-theoretic cost associated with discretization in terms of the expected Kullback-Leibler divergence between nearby continuous distributions characterized by the discretization geometry.
In the original UCN construction, discretization is implemented by transporting the probability mass within each cell to its geometric centroid. The resulting transport-information correspondence, however, is derived under specific conditions involving the cell geometry, spatial coverage, and the vanishing-cell-scale limit. In this work, we generalize the local-map discretization framework by allowing an arbitrary representative point for each cell and establish an exact decomposition of the associated transport cost into contributions from the conditional covariance of the probability distribution and the displacement of the representative point from its probability-weighted centroid. 
Consequently, the resulting transport cost is exactly represented by the expectation of the associated Fisher-induced quadratic form, without requiring uniform cell geometry, complete coverage of the sample space, or an infinitesimal cell-scale limit. The proposed construction thus establishes an exact transport-information correspondence for finite-scale discretization using arbitrary collections of bounded sets.

}
\end{abstract}

\maketitle

\section{Introduction}
The representation change of continuous probability distributions into their discrete counterparts is a fundamental operation in physics and many other areas of science and engineering. Such discretization is often performed by partitioning the sample space into a collection of cells and assigning the probability mass within each cell to a representative point. Although this procedure provides a practical means of reducing a continuous representation to a discrete one, it also introduces a transformation on the sample space. Understanding the structure induced by this transformation therefore requires consideration not only of the statistical distributions themselves but also of the geometric operation through which their representations are changed.

Information geometry\cite{IG} provides a framework for describing the distinguishability of probability distributions through the geometric structure of a statistical manifold. In particular, the Fisher metric characterizes the local distinguishability of nearby distributions and appears as the leading-order quadratic form of the Kullback-Leibler (KL) divergence.\cite{KL} However, information geometry alone does not directly specify the geometric operation performed on the sample space when a continuous distribution is discretized. Consequently, a connection between the cost associated with such an operation and the information-geometric structure of the underlying probability family is not generally guaranteed.

In our recent work, we introduced the concept of unavoidable canonical nonlinearity (UCN),\cite{UCN1, UCN2} establishing a non-arbitrary correspondence between the transport cost induced by discretization and the information-geometric indistinguishability of nearby continuous probability distributions. In this framework, discretization is formulated as a local-map (LM) operation in which the probability mass within each cell is transported to its geometric centroid. By considering the vanishing-cell-scale limit, the resulting transport cost corresponds to the statistical indistinguishability as expected KL divergence on continuous statistical manifold. 

The transport-information correspondence established in the original UCN construction is derived under specific conditions imposed on the discretization scheme. In particular, the cells are constructed from translated copies of a common geometric shape and scale, and the resulting partition covers the entire sample space. Furthermore, the correspondence with the information-geometric quadratic form is obtained in the vanishing-cell-scale limit. These conditions restrict the range of discretization schemes for which the correspondence has been established. In particular, the transport cost associated with a finite cell generally contains contributions arising not only from the conditional spread of the probability distribution within the cell but also from the displacement of the representative point relative to its probability-weighted centroid. This observation motivates a more general examination of the role of representative-point selection in the transport-information correspondence.

In this work, we generalize the local-map (LM) discretization framework by allowing an arbitrary representative point for each cell and examining the resulting transport cost. As in the original UCN construction, we consider a collection of bounded sets as discretization cells in the sample space and define discretization as an operation that transports the probability mass within each cell to its representative point. We show that the transport cost associated with each cell admits an exact decomposition into two contributions: one determined by the conditional covariance of the probability distribution within the cell, and the other by the displacement of the representative point from the probability-weighted centroid. This decomposition provides a general characterization of the transport cost for arbitrary representative-point choices.

We show that the resulting transport cost is given by the expectation of the expected Fisher-induced quadratic form with respect to the parameter fluctuation induced by the cell geometry. The resulting correspondence is exact at finite cell scales and does not require uniform cell geometry or complete coverage of the sample space. The exact equality concerns the Fisher-induced quadratic form, while its interpretation in terms of KL-based indistinguishability remains based on the local expansion of the KL divergence.

The proposed framework thus extends the transport-information correspondence to finite-scale discretization using arbitrary collections of bounded sets. By separating the contribution of within-cell statistical fluctuations from that of representative-point displacement, it clarifies the general structure of the transport cost induced by an LM operation. This establishes a generalized construction of the UCN transport-information correspondence beyond the uniform infinitesimal discretization setting.

\section{Concept and Construction}

\subsection{Original UCN Framework and Its Discretization Scheme}
The unavoidable canonical nonlinearity (UCN) framework\cite{UCN1} provides a connection between the geometric cost induced by discretization on the sample space and the information-geometric indistinguishability of nearby continuous probability distributions. The central idea is to primarily quantify the effect of discretization as a transport operation, and then examine how the resulting transport cost appears on a statistical manifold.

Let $P_\xi(x)$ be a continuous probability distribution on an $f$-dimensional sample space, where $\xi=(\xi_1,\ldots,\xi_f)^T$ denotes a parameter vector associated with the sample-space coordinates. We consider a discretization of the sample space into cells $\{V_k\}$ as any bounded set, each represented by a point $q'_k$ as the geometric center of $V_{k}$. The discretized probability assigned to the representative point is given by
\begin{eqnarray}
P_d(q'_k)=\int_{V_k}P_\xi(x)\,dx.
\end{eqnarray}
In the original UCN construction, the discretization is implemented through a local-map (LM) operation: the probability mass contained in each cell is transported to its representative point. The corresponding squared transport cost based on the 2-Wasserstein framework\cite{WS} is defined as
\begin{eqnarray}
\widetilde W_2^2(P_\xi,P_d)
=
\sum_k\int_{V_k}
(x-q'_k)^T\Omega(x-q'_k)
P_\xi(x)\,dx,
\label{eq:UCN_transport}
\end{eqnarray}
where $\Omega$ is a positive-definite metric on the sample space. This quantity represents the transport cost associated with the specified LM map, rather than the exact 2-Wasserstein distance obtained by minimizing over all admissible transport plans.

The choice of $\Omega$ is constrained by requiring consistency with the local information-geometric structure of the continuous distribution, leading to a Fisher metric for a selected parameter $\xi$ on statistical submanifold. Specifically, we consider a statistical family whose parameter  $\xi$ is directly identifiable with the coordinates of the underlying sample space. In addition, the parametrization is required to transform covariantly under an arbitrary invertible linear transformation of the sample-space coordinates:
\begin{eqnarray}
x\mapsto Ax
\quad\Longrightarrow\quad
\xi\mapsto A\xi,
\qquad A\in GL(f).
\label{eq:GL_covariance}
\end{eqnarray}
The Kullback-Leibler (KL) divergence between nearby members of the family then has the local expansion
\begin{eqnarray}
D(P_{\xi+\delta\xi}:P_\xi)
=
\frac{1}{2}\delta\xi^T\Omega\delta\xi
+o(\|\delta\xi\|^2).
\label{eq:UCN_KL_local}
\end{eqnarray}
The original UCN scheme employs a regular discretization in which the cells are translated copies of the same shape and size. For example, in the isotropic case, the sample space is partitioned into $f$-dimensional hypercubic cells of side length $d$, with each cell represented by its geometric center. The second-moment matrix of a cell, measured relative to its representative point, is then
\begin{eqnarray}
M
=
\frac{1}{|V_k|}
\int_{V_k}
(x-q'_k)(x-q'_k)^T\,dx
=
\frac{d^2}{12}I_f.
\label{eq:UCN_cell_moment}
\end{eqnarray}
More generally, $M$ is determined by the geometry of the common cell shape. In the vanishing discretization limit $d\to0$, the LM-induced transport cost can be expressed, to leading order, in terms of this cell second moment and the Fisher metric:
\begin{eqnarray}
\widetilde W_2^2(P_\xi,P_d)
=
\operatorname{Tr}(M\Omega).
\label{eq:UCN_transport_leading}
\end{eqnarray}
On the information-geometric side, the parameter variation $\delta\xi$ is associated with the discretization scale and extended to a random vector $\rho(\delta\xi)$ whose covariance is proportional to the cell second moment. At $d\to 0$ with the condition of $\operatorname{Cov}\left[\rho\left(\delta\xi\right)\right]=M$, Eq. (\ref{eq:UCN_KL_local}) gives
\begin{eqnarray}
2\mathbb{E}_{\rho}
\left[
D(P_{\xi+\delta\xi}:P_\xi)
\right]
=
\operatorname{Tr}(M\Omega).
\end{eqnarray}
Consequently, the original UCN construction establishes the leading-order correspondence
\begin{eqnarray}
\boxed{
\widetilde W_2^2(P_\xi,P_d)
=
2\mathbb{E}_{\rho}
\left[
D(P_{\xi+\delta\xi}:P_\xi)
\right]
}
\qquad (d\to0).
\label{eq:UCN_correspondence}
\end{eqnarray}
This relation provides an information-geometric interpretation of the transport cost induced by discretization. Namely, the geometric distortion caused by projecting probability mass onto discrete representatives can be represented, to leading order, by the expected indistinguishability of infinitesimally displaced continuous distributions. The correspondence is therefore not simply an algebraic equality between two independently specified quadratic forms: it connects an extrinsic operation defined on the sample space to the intrinsic statistical geometry of the underlying continuous family.

It is important, however, to distinguish the conditions of this original construction from the general scope of the UCN viewpoint. The specific discretization scheme and the transport-information correspondence derived above rely on the following conditions: (i) the cells are translated copies of a common shape, with a common discretization scale; (ii) the cells constitute a tessellation of the entire sample space, as in the partition of $\mathbb{R}^f$; and (iii) the correspondence is obtained in the vanishing-cell-scale limit, $d\to0$. Thus, the expression in Eq. (\ref{eq:UCN_correspondence}) is a leading-order result for this particular discretization scheme, rather than a general finite-scale identity for arbitrary cell configurations.
These conditions become relevant when considering discretization at finite cell sizes. In that setting, the transport cost can depend not only on the shape and size of individual cells, but also on how the entire collection of cells is positioned relative to the probability distribution. In particular, the choice of representative point within each cell may affect the resulting transport cost when the probability distribution is nonuniform.

\subsection{Exact Transport-Information Cost}
In the local-map (LM) discretization scheme, the probability mass within each cell is transported to a representative point assigned to that cell. The choice of this representative point determines the transport operation on the sample space and can therefore affect the resulting transport cost. In the original UCN construction, the representative point is chosen as the geometric centroid of the cell, independently of the probability distribution being discretized. More generally, however, the representative point can be selected arbitrarily, and its role can be examined through an exact decomposition of the corresponding transport cost.

To clarify this structure, let \(S_i\) be an arbitrary bounded set with probability mass

\begin{eqnarray}
V_i=\int_{S_i}P(x)\,dx.
\end{eqnarray}
We introduce the probability-weighted centroid of the distribution restricted to this cell as
\begin{eqnarray}
\bar{x}_i
=
\frac{1}{V_i}
\int_{S_i}xP(x)\,dx,
\end{eqnarray}
and denote its conditional covariance matrix by
\begin{eqnarray}
M_i
=
\frac{1}{V_i}
\int_{S_i}
(x-\bar{x}_i)(x-\bar{x}_i)^T
P(x)\,dx.
\end{eqnarray}
Let \(r_i\) be an arbitrary representative point to which the probability mass within \(S_i\) is transported. Using a common positive-definite transport metric \(\Omega\), the corresponding squared transport cost is
\begin{eqnarray}
C_i(r_i)
=
\int_{S_i}
(x-r_i)^T\Omega(x-r_i)P(x)\,dx.
\end{eqnarray}
By decomposing the displacement as \(x-r_i=(x-\bar{x}_i)+(\bar{x}_i-r_i)\) and using the definition of the probability-weighted centroid, we obtain the exact relation
\begin{eqnarray}
C_i(r_i)
=
V_i\operatorname{Tr}(M_i\Omega)
+
V_i(\bar{x}_i-r_i)^T
\Omega
(\bar{x}_i-r_i).
\label{eq:representative_point_decomposition}
\end{eqnarray}
The two contributions in Eq. (\ref{eq:representative_point_decomposition}) can be expressed in a unified matrix form. Specifically, defining the representative-point displacement as \(x'_i=\bar{x}_i-r_i\), we introduce
\begin{eqnarray}
M'_i=x'_i{x'_i}^T,
\qquad
L_i=M_i+M'_i.
\end{eqnarray}
Equation (\ref{eq:representative_point_decomposition}) can then be rewritten as
\begin{eqnarray}
C_i(r_i)
=
V_i\operatorname{Tr}(M_i\Omega)
+
V_i\operatorname{Tr}(M'_i\Omega)
=
V_i\operatorname{Tr}(L_i\Omega).
\label{eq:general_transport_cost}
\end{eqnarray}
Here, \(M_i\) characterizes the intrinsic spread of the conditional probability distribution within the cell, whereas \(M'_i\) represents the contribution arising from the displacement of the representative point relative to the probability-weighted centroid. Thus, these two matrices describe distinct aspects of the discretization: the former reflects the spatial distribution of probability mass within the cell, while the latter accounts for the offset introduced by the representative-point selection. Their sum \(L_i\) is the second-moment matrix of the conditional probability distribution measured relative to the representative point \(r_i\), namely,
\begin{eqnarray}
L_i
=
\frac{1}{V_i}
\int_{S_i}
(x-r_i)(x-r_i)^T P(x)\,dx.
\end{eqnarray}
This decomposition provides a general description of the transport cost induced by discretization, independently of how the representative point is selected. In particular, when the representative point is chosen as the geometric centroid of the cell, the second contribution reflects the mismatch between the geometry-defined representative point and the actual distribution of probability mass within the cell. Consequently, even for a fixed cell geometry, the transport cost can depend on the spatial distribution of probability mass relative to the cell.

We emphasize that the use of geometric centroids is not intrinsically inappropriate. It provides a natural discretization when the representative points are intended to reflect the geometry of the sample-space partition itself. Nevertheless, Eq. (\ref{eq:general_transport_cost}) shows that the transport cost generally contains both the contribution from the conditional covariance and that from the representative-point displacement. The latter vanishes when the representative point coincides with the probability-weighted centroid. Furthermore, since \(\Omega\) is positive definite, the probability-weighted centroid uniquely minimizes the transport cost among all possible representative points for a given cell.


We now derive the transport cost induced by the distribution-adaptive discretization scheme and establish its correspondence with the Fisher-induced quadratic form. Let $\Omega$ denote the Fisher metric of the underlying continuous probability family, as described in Sec. II.A. In the proposed LM operation, the probability mass within each cell $S_i$ is transported to an arbitrary  representative point $\bar{r}_i$. In this case, again, the corresponding transport cost for the $i$-th cell is given by
\begin{eqnarray}
C_i
= V_{i} \textrm{Tr}\left(L_{i}\Omega\right).
\end{eqnarray}

We next express this transport cost in terms of the Fisher-induced quadratic form. 
In a similar fashion to the UCN construction, we introduce probability distribution $\rho_i(\delta\xi)$ on the parameter space whose second moment is chosen to coincide with $L_i$, namely,
\begin{eqnarray}
\mathbb{E}_{\rho_{i}}\left[\delta\xi\delta\xi^{\textrm{T}}\right] = L_{i}
\end{eqnarray}
The Fisher metric determines the local quadratic form associated with the distinguishability of nearby distributions in the statistical manifold. We define this quadratic form as
\begin{eqnarray}
Q(\delta\xi)
=
\frac{1}{2}
\delta\xi^T\Omega\delta\xi,
\end{eqnarray}
where $\delta\xi$ denotes a displacement in the parameter space of the continuous probability family. Then the expectation of this quadratic form with respect to $\rho_i$ is given by
\begin{eqnarray}
2\mathbb{E}_{\rho_i}[Q(\delta \xi)]
&=&
\mathbb{E}_{\rho_i}
[\delta \xi^T\Omega\delta \xi]
\nonumber\\
&=&
\operatorname{Tr}(L_i\Omega).
\end{eqnarray}
Combining this relation with the expression for $C_i$, we obtain
\begin{eqnarray}
C_i
=
2V_i\mathbb{E}_{\rho_i}[Q(\delta \xi)].
\end{eqnarray}
This equality establishes an exact correspondence between the transport cost associated with each cell and the expected Fisher-induced quadratic form. Importantly, the equality does not require the cell size to approach zero or the cell geometry to be uniform. 

The total transport cost over the collection of cells is obtained by summing the individual contributions:
\begin{eqnarray}
C
&=&
\sum_i C_i
\nonumber\\
&=&
\sum_i V_i\operatorname{Tr}(L_i\Omega)
\nonumber\\
&=&
2\sum_i V_i
\mathbb{E}_{\rho_i}[Q(\delta \xi)].
\end{eqnarray}
When we define
\begin{eqnarray}
L = \sum_{i} V_{i} L_{i}, \,\,\, V = \sum_{i} V_{i},
\end{eqnarray}
we can finally obtain the exact transport-information correspondence for discretization:
\begin{eqnarray}
\boxed{
C = \textrm{Tr}\left(L\Omega\right) =2\mathbb{E}_{\rho}\left[Q\left(\delta\xi\right)\right]
}
\end{eqnarray}
where
\begin{eqnarray}
\mathbb{E}_{\rho}\left[\delta\xi \delta\xi^{\textrm{T}}\right] = L.	
\end{eqnarray}

This expression establishes for an arbitrary collection of bounded sets. The cells may have different shapes and sizes, and their union need not cover the entire sample space. In the latter case, $C_{\mathrm{total}}$ represents the transport cost associated only with the probability mass contained in the selected cells; probability mass outside their union is not included in the discretization operation considered here.

We distinguish the exact correspondence derived above from the local information-geometric expansion discussed in Sec. II.A. The equality between the transport cost and the expected Fisher-induced quadratic form is exact for the proposed discretization scheme. By contrast, identifying $Q(\delta x)$ with the Kullback--Leibler divergence between nearby distributions is justified only to leading order in the local parameter variations, through the Fisher metric. Accordingly, the present construction establishes an exact finite-scale correspondence at the level of the Fisher-induced quadratic form, while its interpretation in terms of KL-based statistical indistinguishability retains the local character of information geometry.

The present formulation can be further extended to discretization schemes in which the probability mass within a single cell is distributed among multiple representative points, rather than being transported entirely to a single point. Specifically, one may assign a nonnegative scalar fraction of the probability density within each cell to each representative point, provided that these fractions are independent of the position within the cell. The resulting transport cost is obtained by summing the contributions associated with the individual representative points, and the corresponding second-moment matrix is given by the weighted sum of the matrices defined relative to each representative point. Consequently, the same exact transport-information correspondence derived above remains valid without modification. Furthermore, the sum of the assigned fractions need not equal unity, allowing the framework to describe operations in which only a portion of the probability mass within a cell is transported to discrete representatives. A brief derivation of this extension is provided in Appendix.

\subsection{Extension to Discrete Distributions}
The distribution-adaptive discretization framework introduced above can also be formulated for discrete probability distributions. In this case, the sample space consists of discrete support points, and a collection of cells is defined by grouping these points. Let \(p_j\) denote the probability assigned to the support point \(x_j\), and let \(S_i\) denote the \(i\)-th cell. The probability mass contained in each cell is given by
\begin{eqnarray}
V_i=\sum_{j\in S_i}p_j.
\end{eqnarray}
For each cell with \(V_i>0\), we introduce an arbitrary representative point \(r_i\) to which the probability mass within the cell is transported. Using a positive-definite transport metric \(\Omega\) defined on the support-coordinate space, the corresponding cell-wise transport cost is
\begin{eqnarray}
C_i(r_i)
=
\sum_{j\in S_i}
p_j(x_j-r_i)^T
\Omega
(x_j-r_i).
\end{eqnarray}
By decomposing the displacement into contributions from the conditional spread within the cell and the offset of the representative point from the probability-weighted centroid, we obtain
\begin{eqnarray}
C_i(r_i)
=
V_i\operatorname{Tr}(M_i\Omega)
+
V_i\operatorname{Tr}(M'_i\Omega).
\end{eqnarray}
Thus, the same exact decomposition as in the continuous case holds for discrete distributions. In particular, choosing the probability-weighted centroid \(r_i=\bar{x}_i\) eliminates the contribution associated with the representative-point displacement, yielding
\begin{eqnarray}
C_i
=
V_i\operatorname{Tr}(M_i\Omega).
\end{eqnarray}

To establish a transport-information correspondence for discrete distributions, the transport metric \(\Omega\) must be identified consistently with the Fisher metric of the statistical model under consideration. A natural setting in which such a metric is available is a discrete exponential family. Consider a family of the form
\begin{eqnarray}
p_j(\xi)
=
p_j^0
\exp\left(
\xi^T x_j-\psi(\xi)
\right),
\end{eqnarray}
where \(x_j\) denotes the sufficient statistic associated with the support point, \(\xi\) is the natural parameter, and \(\psi(\xi)\) is the log-partition function. We introduce the expectation parameter
\begin{eqnarray}
\eta
:=
\mathbb{E}_{p_\xi}[x]
=
\nabla_\xi\psi(\xi).
\end{eqnarray}
For a regular minimal exponential family, the natural and expectation parameters are locally related through
\begin{eqnarray}
\frac{\partial\eta}{\partial\xi}
=
\nabla_\xi^2\psi(\xi)
=
\operatorname{Cov}_{p_\xi}(x).
\end{eqnarray}
Here, \(\eta\) has the same component structure and units as the support coordinate \(x\), making it suitable for describing local variations in the support-coordinate space.

The Fisher metric in the natural-parameter coordinates is given by
\begin{eqnarray}
\Omega_{\mathrm{F}}(\xi)
=
\nabla_\xi^2\psi(\xi)
=
\operatorname{Cov}_{p_\xi}(x).
\end{eqnarray}
Transforming this metric to the expectation-parameter coordinates, we obtain
\begin{eqnarray}
\Omega_{\mathrm{F}}(\eta)
=
\operatorname{Cov}_{p_{\xi(\eta)}}(x)^{-1}
\end{eqnarray}
where \(\xi(\eta)\) denotes the natural parameter corresponding to the expectation parameter \(\eta\). Thus, the Fisher metric in the expectation-parameter coordinates is given by the inverse of the covariance matrix of the support coordinates under the full probability distribution. This covariance matrix characterizes the global statistical spread of the original distribution and is distinct from the cell-wise covariance matrices \(M_i\) appearing in the discretization-induced transport cost. In particular, its definition does not require the origin of the support-coordinate space to coincide with the mean of the distribution, since the covariance matrix is invariant under translations of the support coordinates.

This choice of parameterization resolves the coordinate distinction between the sample-space operation and the information-geometric description. In contrast to the natural parameter \(\xi\), which is dual to the support coordinate \(x\), the expectation parameter \(\eta\) belongs to the same coordinate space as \(x\). Consequently, the Fisher metric \(\Omega_{\mathrm{F}}(\eta)\) can be consistently considered as a metric on the support-coordinate space, allowing it to be used as the transport metric in the discretization framework. The corresponding local information-geometric quadratic form is
\begin{eqnarray}
Q(\delta\eta)
=
\frac{1}{2}
\delta\eta^T
\Omega_{\mathrm{F}}(\eta)
\delta\eta,
\end{eqnarray}
which gives the second-order approximation of the Kullback--Leibler divergence between neighboring distributions parameterized by \(\eta\).

Accordingly, identifying the transport metric with \(\Omega_{\mathrm{F}}(\eta)\) establishes a transport-information correspondence in which the sample-space transport cost and the local statistical distinguishability are expressed in the same coordinate space. Once this identification is specified, the same covariance-metric contraction provides the corresponding Fisher-induced quadratic structure for the discrete discretization operation. This extension indicates that the distribution-adaptive construction is not restricted to continuous probability distributions, while also clarifying the additional coordinate considerations required for its application to discrete statistical models.t

We finally note that the present formulation does not require the cells to be convex and can therefore be applied to arbitrary bounded measurable sets, provided that the relevant probability moments are finite. It can also be extended to overlapping cells by introducing an appropriate allocation scheme for probability mass in the overlapping regions. Such an extension would further allow the allocation scheme itself to be optimized with respect to the total transport cost. Such generalized treatments are briefly discussed in Appedix.

 \section{ Conclusions}
In this work, we have generalized the transport-information correspondence developed in the framework of unavoidable canonical nonlinearity (UCN) by reconsidering the role of representative-point selection in local-map (LM) discretization. We first established that, for an arbitrary representative point in each cell, the induced transport cost admits an exact decomposition into two contributions: one arising from the conditional covariance of the probability distribution within the cell, and the other from the displacement of the representative point relative to its probability-weighted centroid. This result provides a general characterization of the transport cost induced by LM discretization, independently of the particular representative-point selection.

Based on this decomposition, we derive an exact expression for the cell-wise transport cost in terms of the cell probability mass, the conditional covariance matrix, and the Fisher metric of the underlying continuous probability family. Furthermore, we established an exact equality between the transport cost and the expected Fisher-induced quadratic form. Thus, the correspondence arises directly from the statistical structure of the discretization operation.

An important consequence of this construction is that the transport-information correspondence does not require uniform cell geometry, complete coverage of the sample space, or a vanishing-cell-scale limit. It holds for arbitrary collections of bounded sets, including cells with different shapes and sizes. We emphasize that the exact correspondence is established at the level of the Fisher-induced quadratic form, whereas its interpretation in terms of KL-based statistical indistinguishability remains based on the local expansion of the KL divergence. 

The present results clarify how the representative-point selection determines the transport cost induced by discretization and how  an exact finite-scale connection with information geometry is established. This extends the UCN transport-information correspondence beyond the uniform infinitesimal discretization setting and provides a general construction for relating sample-space operations to the information-geometric structure of probability distributions.

\section{Acknowledgement}
This work was supported by Research Grant from Hitachi Metals$\cdot$Materials Science Foundation.

\section*{Appendix : Extension to Multiple Representative Points}
We consider an extension of the LM discretization scheme in which the probability mass within each cell $S_i$ is distributed among multiple representative points $\{r_{i\alpha}\}_{\alpha}$. Let $\lambda_{i\alpha}$ be the fraction of the probability density assigned to the representative point $r_{i\alpha}$, where
\begin{eqnarray}
\lambda_{i\alpha}\geq 0,
\qquad
\sum_{\alpha}\lambda_{i\alpha}\leq 1.
\end{eqnarray}
Here, the fractions are assumed to be independent of the position $x$ within the cell. Thus, the fraction $\lambda_{i\alpha}P(x)$ of the probability density at each point $x\in S_i$ is transported to $r_{i\alpha}$. When the sum of the fractions is unity, the entire probability mass within the cell is transported, whereas a sum smaller than unity corresponds to an operation in which only a portion of the cell's probability mass is included.

For each representative point, we define the corresponding second-moment matrix as
\begin{eqnarray}
L_{i\alpha}
=
\frac{1}{V_i}
\int_{S_i}
(x-r_{i\alpha})(x-r_{i\alpha})^T P(x)\,dx.
\end{eqnarray}
Using the decomposition introduced in Sec. II.B, this matrix can be expressed as
\begin{eqnarray}
L_{i\alpha}=M_i+M'_{i\alpha},
\end{eqnarray}
where
\begin{eqnarray}
M_i 
&=&
\frac{1}{V_i}
\int_{S_i}
(x-\bar{x}_i)(x-\bar{x}_i)^T
P(x)\,dx  \nonumber \\
M'_{i\alpha}
&=&
(\bar{x}_i-r_{i\alpha})
(\bar{x}_i-r_{i\alpha})^T.
\end{eqnarray}

The total transport cost associated with the $i$-th cell is then given by the sum of the contributions from all representative points:
\begin{eqnarray}
C_i^{\mathrm{mt}}
&=&
\sum_{\alpha}\lambda_{i\alpha}
\int_{S_i}
(x-r_{i\alpha})^T
\Omega
(x-r_{i\alpha})P(x)\,dx
\\
&=&
V_i\operatorname{Tr}
\left(
L_i^{\mathrm{mt}}\Omega
\right),
\end{eqnarray}
where the effective second-moment matrix is defined as
\begin{eqnarray}
L_i^{\mathrm{mt}}
&=&
\sum_{\alpha}\lambda_{i\alpha}L_{i\alpha}
\\
&=&
\left(\sum_{\alpha}\lambda_{i\alpha}\right)M_i
+
\sum_{\alpha}\lambda_{i\alpha}M'_{i\alpha}.
\end{eqnarray}
Since $L_i^{\mathrm{mt}}$ is positive semidefinite, we can introduce probability distribution $\rho_i^{\mathrm{mt}}(\delta\xi)$ whose second moment is given by
\begin{eqnarray}
E_{\rho_i^{\mathrm{mt}}}
[\delta\xi\delta\xi^T]
=
L_i^{\mathrm{mt}}.
\end{eqnarray}
Using the Fisher-induced quadratic form $Q(\delta\xi)=\frac{1}{2}\delta\xi^T\Omega\delta\xi$, the transport cost can therefore be expressed as
\begin{eqnarray}
C_i^{\mathrm{mt}}
&=&
V_i\operatorname{Tr}
\left(L_i^{\mathrm{mt}}\Omega\right)
\\
&=&
2V_i E_{\rho_i^{\mathrm{mt}}}
[Q(\delta\xi)].
\end{eqnarray}
Thus, the exact transport-information correspondence established in Sec. II.B is preserved when the probability mass within each cell is distributed among multiple representative points with position-independent allocation fractions. The total correspondence follows by summing the cell-wise contributions, in the same manner as in the single-representative-point scheme.

The total transport cost over the collection of cells is obtained by summing the individual contributions:
\begin{eqnarray}
C^{\mathrm{mt}}
&=&
\sum_i C_i^{\mathrm{mt}}
\\
&=&
\sum_i V_i
\operatorname{Tr}
\left(L_i^{\mathrm{mt}}\Omega\right)
\\
&=&
2\sum_i V_i
E_{\rho_i^{\mathrm{mt}}}
[Q(\delta\xi)].
\end{eqnarray}
Defining the total second-moment matrix as
\begin{eqnarray}
L^{\mathrm{mt}}
=
\sum_i V_i L_i^{\mathrm{mt}},
\end{eqnarray}
we obtain
\begin{eqnarray}
C^{\mathrm{mt}}
=
\operatorname{Tr}
\left(L^{\mathrm{mt}}\Omega\right).
\end{eqnarray}
We can then introduce probability distribution $\rho^{\mathrm{mt}}(\delta\xi)$ whose second moment is given by
\begin{eqnarray}
E_{\rho^{\mathrm{mt}}}
[\delta\xi\delta\xi^T]
=
L^{\mathrm{mt}}.
\end{eqnarray}
Consequently, the total transport cost satisfies
\begin{eqnarray}
\boxed{
C^{\mathrm{mt}}
=
\operatorname{Tr}
\left(L^{\mathrm{mt}}\Omega\right)
=
2E_{\rho^{\mathrm{mt}}}
[Q(\delta\xi)].
}
\end{eqnarray}
Thus, the exact transport-information correspondence holds not only for each individual cell but also for the total cost associated with the entire collection of cells, irrespective of the number of representative points assigned to each cell.

We finally note that the present extension can also accommodate discretization schemes involving overlapping cells. For a finite collection of overlapping cells, the union of these cells can be decomposed into mutually disjoint subregions, each characterized by the specific combination of original cells to which it belongs. Each resulting subregion can then be treated as a cell in the present formulation, with the representative points associated with the corresponding original cells considered as candidate representatives. By assigning probability-mass fractions among these representatives within each subregion, the same transport-information correspondence can be applied without double-counting the probability mass in overlapping regions. Thus, overlapping-cell schemes can also be incorporated into the present framework through an appropriate refinement of the cell structure. Furthermore, the above formulation based on disjoint subregions can encompass schemes in which the original overlapping cells are treated independently, without explicitly subdividing their overlap regions. Specifically, any transport plan defined on the original cells can be reproduced by imposing appropriate consistency constraints on the allocation fractions across the resulting subregions. The refined formulation therefore provides a more general class of admissible transport plans, while retaining the original schemes as a special case.

\end{document}